\documentclass[fp,twocolumn]{jpsj3}  
\usepackage{graphicx}
\usepackage{color}
\usepackage{hhline}
\usepackage{mathrsfs}
\usepackage{dcolumn}
\usepackage{bm}
\usepackage{multirow}
\usepackage{booktabs}
\usepackage{afterpage}
\usepackage{amsmath}
\usepackage{braket}

\begin{document}

\title{DMFT study on the electron-hole asymmetry \\ of the electron correlation strength in the high $T_{\rm c}$ cuprates}

\author{Ryota Mizuno\thanks{mizuno@presto.phys.sci.osaka-u.ac.jp}, Masayuki Ochi, Kazuhiko Kuroki}

\inst{ Department of Physics, Osaka University, 1-1 Machikaneyama, Toyonaka, Osaka 560-0043, Japan}

\date{\today}
\abst{
Recent experiments revealed a striking asymmetry in the phase diagram of the high temperature cuprate superconductors.
The correlation effect seems strong in the hole-doped systems and weak in the electron-doped systems.
On the other hand, 
a recent theoretical study shows that the interaction strengths (the Hubbard $U$) are comparable in these  systems. 
Therefore, 
it is difficult to explain this asymmetry by their interaction strengths.
Given this background, 
we analyze the one-particle spectrum of a single band model of a cuprate superconductor  near the Fermi level using the dynamical mean field theory. We find the difference in the ``visibility'' of the strong correlation effect  between the hole- and electron-doped systems. 
This can explain the electron-hole asymmetry of the correlation strength without introducing the difference in the interaction strength.
}

\maketitle

\section{Introduction}
Over thirty years have passed since the high temperature cuprate superconductors were discovered in 1986. Despite a huge number of studies,
there still remain various unsolved problems in the study of this family of unconventional superconductors.
The remarkable difference in the phase diagram of the cuprates between hole- and electron-doped materials is one of these unsolved issues.
The mother compounds of the high temperature cuprate superconductors are usually insulating,
and doping carriers, holes or electrons, induces superconductivity.
It has been believed that the mother compounds of cuprates are Mott insulators due to their strong electron correlation.
However,
recent experimental studies have shown that,
in the electron-doped systems, 
the electron correlation strength is not so strong that even very lightly doped samples exhibit superconductivity if they are properly annealed and the antiferromagnetism is suppressed. Some of these studies suggest that superconductivity persists even with no electron doping\cite{PhysRevLett.74.4927,Tsukada2005427,2}, 
while others show presence of antiferromagnetic long range order in the lightly doped regime\cite{Saadaoui20150121online,  ft_1}, 
so that the issue is still controversial.
It was also shown experimentally that the ``pseudo gap''\cite{PhysRevLett.63.1700, PhysRevLett.62.1193, norman1998destruction,PhysRevB.63.024504,RevModPhys.82.2421}, 
a prominent feature of the cuprates,
disappears if the antiferromagnetism is suppressed\cite{d845815e89a94214b3c7398b7a34611f}.
This is consistent with a theoretical study, which shows that the pseudo gap in the electron-doped case originates from the antiferromagnetism\cite{PhysRevB.76.012502}.
By contrast,
in the hole-doped systems,
the pseudo gap exists even in the doping regime far away from the antiferromagnetic phase.
Therefore, 
the formation mechanism of the pseudo gap in the hole-doped systems appears to be different from that in the electron-doped systems. 
A plausible explanation for the appearance of the pseudo gap in the hole-doped materials is the  Mott insulating state, or the Mottness, of their mother compounds\cite{PhysRevLett.86.139}.

\begin{figure}
\centering
\includegraphics[width=7.5cm,clip]{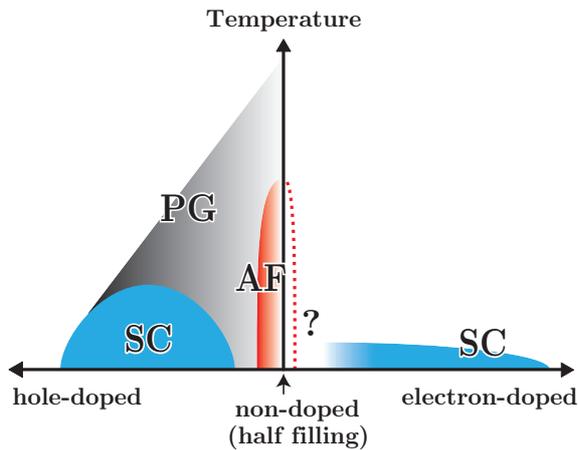} 
\caption{(Color online) A schematic phase diagram of the high temperature cuprate superconductors: 
The pseudo gap exists widely over the antiferromagnetic phase and the dome-shaped superconducting phase in the hole-doped side, 
while $T_c$ of the superconducting phase varies monotonically in the electron-doped side.
The realization of superconductivity or antiferromagnetism in the mother compound, or in the very vicinity of it,  is still  controversial, especially for the electron-doped-type materials.
}
\label{fig:2017-04-03-14-58}
\end{figure}

The above mentioned 
electron-hole asymmetry of the cuprates  is summarized in the phase diagram shown in Fig. \ref{fig:2017-04-03-14-58}.
There is a theoretical calculation that explains the asymmetry of the superconducting phase (dome-shaped vs. monotonic) by treating the O-$2p$ orbitals properly\cite{Ogura}.
 On the other hand, 
the electron-hole asymmetry of the presence/absence of the pseudo gap phase is still an open question. It is important to note that this asymmetry occurs in a rather {\it abrupt} manner, i.e., there is a large difference between the hole and the electron underdoped regimes. 
Since the origin of the pseudo gap in the hole-doped regime is most likely to be the strong electron correlation, one may consider that the origin of this asymmetry lies in the difference in the effective electron-electron interaction strength between the hole- and electron-doped materials\cite{5}.
In fact, the electron-doped cuprates have been considered to exhibit weak correlation compared to a hole-doped cuprate La$_2$CuO$_4$ due to the small Cu $3d$-O $2p$ level offset. A material with a small $d$-$p$ offset has small on-site Hubbard $U$ when mapped to an effective single band model. The difference in the $d$-$p$ offset and hence the on-site $U$ rises because the electron-doped cuprates, with the T'-type structure, have no apical oxygen, while La$_2$CuO$_4$ having the T-type structure has apical oxygens at close distance to the copper atom\cite{ea8b98346b67403f91716b92319ddfe3,PhysRevB.82.125107,PhysRevB.79.134522,doi:10.7566/JPSJ.82.063713}.
However, 
a recent first principles estimation\cite{Jang.Seung.Woo.1} has revealed that La$_2$CuO$_4$ is actually an exception that has very large $d$-$p$ level offset, while many of the hole-doped cuprates have $d$-$p$ offset values and hence the on-site $U$ comparable to  those in the T'-type electron doped systems. 
A typical example with a moderate $d$-$p$ offset and $U$ is ${\rm HgBa_{2}CuO_{4}}$, in which $T_c$ is very high ($\sim 100$K) and a pseudo gap is observed\cite{PhysRevB.63.024504}.
Therefore,
it is difficult to explain the electron-hole asymmetry in the diagram by their interaction strength.

Given this background,
we analyze the one-particle spectrum of a single band model of a cuprate superconductor near the Fermi level using dynamical mean field theory (DMFT)\cite{RevModPhys.68.13} with two kinds of impurity solvers : iterated perturbation theory (IPT)\cite{doi:10.1143/PTPS.46.244, doi:10.1143/PTP.53.970, doi:10.1143/PTP.53.1286, Yamada4, PhysRevB.45.6479, PhysRevLett.77.131, modified_IPT_Potthoff,PhysRevB.86.085133}
and continuous time quantum Monte Carlo (CT-QMC)\cite{PhysRevB.76.235123,PhysRevB.72.035122, Rubtsov2004,PhysRevB.89.195146,PhysRevB.80.195111,PhysRevB.83.075122} methods. 
We adopt a model Hamiltonian derived from the first principles band calculation of ${\rm HgBa_{2}CuO_{4}}$ since the realistic shape of the density of states (DOS) is important in the present analysis. We use the same model for both the electron and hole doped cases (just vary the electron density), since the band structure of ${\rm Nd_{2}CuO_{4}}$ is very similar to that of ${\rm HgBa_{2}CuO_{4}}$\cite{Ogura}.

We find 
that the electron-hole asymmetry can exist even under common interaction strengths  
between the hole- and the electron-doped systems.
The asymmetry of the spectrum already exists in the non-interacting case, 
but it is drastically enhanced, especially when the interaction is strong enough to make the non-doped case a Mott insulator.

This paper is organized as follows.
In Sec.\ref{sec:Method}, we describe the outlines of DMFT and the solvers.
Computational details and results are given in Sec.\ref{sec:Result}.
The conclusion is given in Sec.\ref{sec:Conclusion}.

\section{Method}\label{sec:Method}
\subsection{Dynamical Mean Field Theory} \label{sec:2017-04-15-15-01}

We investigate the one-band Hubbard model
\begin{align}
H =&
-\sum_{ij\sigma} t_{ij} c^{\dagger}_{i\sigma} c_{j\sigma} 
-\mu \sum_{i\sigma}c_{i\sigma}^{\dagger}c_{i\sigma}
+ U \sum_{i} n_{i\uparrow}n_{i\downarrow},
\label{eq:2017-04-02-00-47}
\end{align}
where $t_{ij}$ is the hopping matrix between sites $i$ and $j$, 
$\mu$ and $U$ are the chemical potential and the Coulomb repulsion, respectively.
$c_{i\sigma}^{(\dagger)}$ is the annihilation (creation) operator for an electron of spin $\sigma$ on site $i$, and $n_{i\sigma} = c_{i\sigma}^{\dagger}c_{i\sigma}$ is the number operator.

If the wave number dependence of the self energy is neglected, 
the Dyson equation for the local Green's function in the Hubbard model is equivalent to  that of the impurity Green's function  in the Anderson model. 
Therefore,
in the dynamical mean  field theory, 
the lattice problem is reduced to solving the impurity model embedded in an effective bath.
This effective bath is determined by the condition
\begin{align}
\Delta(i\omega_{n})
=&
i\omega_{n} + \mu - \Sigma(i\omega_{n}) \nonumber \\
&-
\left[ \sum_{\bm{k}} \dfrac{1}{i\omega_{n} + \mu - \epsilon_{\bm{k}} - \Sigma(i\omega_{n})}\right]^{-1},
\label{eq:2017-04-02-01-24}
\end{align}
where $\omega_{n}$ and $\bm{k}$ are the Fermionic Matsubara frequency and the wave number, respectively. 
$\Delta(i\omega_{n})$ is the bath hybridization function in the Anderson model, and 
$\epsilon_{\bm{k}}$ is the band dispersion in the Hubbard model.
$\Sigma(i\omega_{n})$ is the self energy common to both models. 

In the actual calculation, 
one can solve the problem in the following steps : 
(i) Determine the initial self energy as $\Sigma(i\omega_{n})=0$. 
(ii) Calculate the hybridization function $\Delta(i\omega_{n})$ by Eq. (\ref{eq:2017-04-02-01-24}).
(iii) Solve the impurity problem using $\Delta(i\omega_{n})$ and obtain the new self energy.
(iv) Back to step (ii).
If this self-consistent loop is converged, 
the Green's function in the lattice problem is evaluated as 
\begin{align}
G_{\bm{k}}(i\omega_{n})
=&
\dfrac{1}{i\omega_{n}+\mu-\epsilon_{\bm{k}} - \Sigma(i\omega_{n})}.
\label{eq:2017-04-02-10-29}
\end{align}

\begin{figure}
\centering
\includegraphics[width=8.5cm,clip]{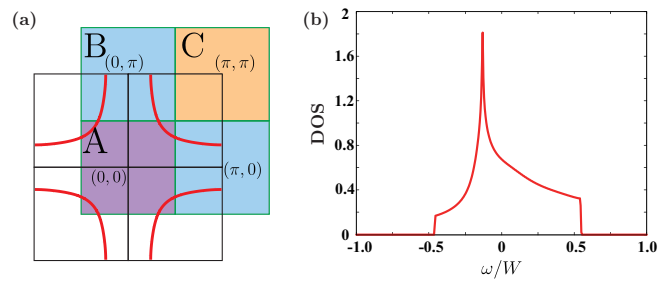} 
\caption{(Color online) (a) Brillouin zone partitioning :
The three inequivalent sectors A, B, and C are described with colors violet, cyan, and yellow, respectively.
The non-interacting Fermi surface of the single-band model of ${\rm HgBa_{2}CuO_{4}}$ at half filling is indicated by the red line. 
(b) The non-interacting density of states of ${\rm HgBa_{2}CuO_{4}}$ at half filling
}
\label{fig:2017-04-02-23-40}
\end{figure}

As mentioned above,
in DMFT, 
the self energy has no $\bm{k}$ dependence.  
One of the extended theories of DMFT to take the $\bm{k}$ dependence into account is the dynamical cluster approximation (DCA)\cite{RevModPhys.77.1027}. 
In this theory, 
$\bm{k}$ dependence of the self energy is partially taken into account by dividing the Brillouin Zone into several sectors.
In this study, 
we use both the single-site (non-extended) DMFT and the DCA.
In  DCA, 
we divide the Brillouin Zone into four sectors as shown in FIG. \ref{fig:2017-04-02-23-40}(a).

\subsection{Impurity Solvers}
To solve the impurity problem,
various numerical methods have been proposed: iterated perturbation theory,  continuous time quantum Monte Carlo,  exact diagonalization (ED)\cite{PhysRevLett.72.1545}, 
non-crossing approximation (NCA)\cite{Kuramoto1983,Kuramoto1984,Kojima1984,Kuramoto1986},  numerical renormalization group (NRG)\cite{RevModPhys.80.395}, etc. 
In the following section, 
we briefly describe IPT and CT-QMC, which we used in this study.

\subsubsection{Iterated perturbation Theory}

In the iterated perturbation theory\cite{doi:10.1143/PTPS.46.244, doi:10.1143/PTP.53.970, doi:10.1143/PTP.53.1286, Yamada4, PhysRevB.45.6479},
one can solve the impurity problem by the second order perturbation theory.
In the electron-hole symmetric case,
it is known that IPT provides a good result\cite{PhysRevB.48.7167,PhysRevLett.70.1666,PhysRevB.49.10181} and reproduces the exact solution in both weak and strong correlation limits\cite{modified_IPT_Potthoff}.
In other cases, however,  the results are not so good. 
To overcome this weakness,
the extended version of IPT for arbitrary filling 
\cite{PhysRevLett.77.131, modified_IPT_Potthoff,PhysRevB.86.085133} 
was proposed. 
In this section, 
we present the formalism of the extended version of IPT that we used in our study.

The self energy is parametrized by 
\begin{align}
\Sigma(i\omega_{n})
=&
U n  + \dfrac{A \Sigma^{(2)}(i\omega_{n})}{1-B\Sigma^{(2)}(i\omega_{n})}
\label{eq:2017-04-02-03-35}
\end{align}
where 
\begin{align}
\Sigma^{(2)}(i\omega_{n})
=&
U^{2}T \sum_{m}G_{0}(i\omega_{n}+i\nu_{m})\chi_{0}(i\nu_{m})
\label{eq:2017-04-02-03-37}\\
\chi_{0}(i\nu_{m})
=&
-T \sum_{l}G_{0}(i\omega_{l})G_{0}(i\omega_{l}+i\nu_{m})
\label{eq:2017-04-02-03-38}
\end{align}
with the zeroth order Green's function
\begin{align}
G_{0}(i\omega_{n})
=&
\dfrac{1}{i\omega_{n} + \mu_{0} - \Delta(i\omega_{n})}
\label{eq:2017-04-02-10-09}
\end{align}
and $\mu_{0}$ and $\nu_{m}$ are the pseudo chemical potential and Bosonic Matsubara frequency, respectively.
The local Green's function in the lattice problem is
\begin{align}
G(i\omega_{n}) 
=& 
\sum_{\bm{k}}G_{\bm{k}}(i\omega_{n})
=&
\sum_{\bm{k}}\dfrac{1}{i\omega_{n} + \mu -\epsilon_{\bm{k}} - \Sigma(i\omega_{n})}.
\label{eq:2017-04-02-10-37}
\end{align}
The constants $A$ and $B$ are determined such that one reproduces the exact solution in the high frequency and the atomic limits: 
\begin{align}
A
=&
\dfrac{n(1-n)}{n_{0}(1-n_{0})}, 
\hspace{5pt}
B
=
\dfrac{\mu_{0}-\mu + (1-n)U}{n_{0}(1-n_{0})U^{2}},
\label{eq:2017-04-02-10-16}
\end{align}
where $n_{0}$ and $n$ are the electron numbers evaluated from $G_{0}(i\omega_{n})$ and $G(i\omega_{n})$, respectively.

The chemical potential $\mu$ is determined by fixing $n$ at the input value, while $\mu_{0}$ is still a free parameter. 
Among the various suggested conditions for determining $\mu_{0}$\cite{PhysRevB.33.1814,modified_IPT_Potthoff,Meyer1999225,PhysRevB.86.085133}, 
here we employ the condition $n=n_{0}$.
In this condition, 
the IPT calculation provides a qualitatively good result  except in the case when the Coulomb interaction strength is too large\cite{modified_IPT_Potthoff,PhysRevB.86.085133}.

\subsubsection{Continuous Time Quantum Monte Carlo Method}

Continuous time quantum Monte Carlo method is numerically exact.
Among the several algorithms of CT-QMC, here we adopt CT-INT 
 in this study.
This algorithm,
 developed by Rubtsov $et$ $al$\cite{PhysRevB.72.035122, Rubtsov2004},
  is based on the Monte Carlo summation of all diagrams obtained from the expansion of the partition function in powers of the interaction $U$.
We describe only a brief outline of CT-INT in this section. 
For efficient computation, 
we also employ the submatrix update algorithm\cite{PhysRevB.80.195111,PhysRevB.83.075122} extended to CT-INT\cite{PhysRevB.89.195146}.

We divide the impurity Hamiltonian into two parts, $H=H_{0}+H_{\rm i}$. 
$H_{\rm i}$ is the interacting term including $U$.
The partition function is 
\begin{align}
Z
=&
{\rm Tr} \left[ e^{-\beta H_{0}} T_{\tau} \exp\left( -\int_{0}^{\beta} d\tau H_{\rm i}(\tau) \right) \right],
\label{eq:2017-04-04-11-07}
\end{align}
where $A(\tau) = e^{\tau H_{0}} A e^{-\tau H_{0}}$ is the interaction representation of the operator $A$, and $T_{\tau}$ is the time ordering operator for the imaginary time.
The statistical average for $H_{0}$ is
\begin{align}
\braket{A}_{0} =& \dfrac{1}{Z_{0}} {\rm Tr}[ e^{-\beta H_{0}}A]
,
\hspace{5pt}
Z_{0} = {\rm Tr}e^{-\beta H_{0}}.
\label{eq:2017-04-04-11-07}
\end{align}
We expand the partition function in powers of $H_{\rm i}$ and obtain
\begin{align}
\dfrac{Z}{Z_{0}} =& \sum_{n=0}^{\infty} \int_{0}^{\beta} d\tau_{n} \cdots d\tau_{2} d\tau_{1}
\nonumber \\
& \hspace{5pt} \times
(-1)^{n}\braket{T_{\tau} H_{\rm i}(\tau_{n}) \cdots  H_{\rm i}(\tau_{2}) H_{\rm i}(\tau_{1}) }_{0}.
\label{eq:2017-04-04-11-30}
\end{align}
The statistical average of the operator $A$ for the full Hamiltonian $H$ is described as 
\begin{align}
\braket{A}
=&
 \dfrac{ \sum_{n} \int_{n} d\tau \braket{\hat{P}(q_{n})A(q_{n})}_{0} }{\sum_{n}\int_{n}d\tau P(q_{n})},
\label{eq:2017-04-04-11-37}
\end{align}
where
\begin{align}
P(q_{n}) =& \braket{\hat{P}(q_{n})}_{0} \nonumber \\
=&
(-1)^{n}\braket{T_{\tau} H_{\rm i}(\tau_{n}) \cdots  H_{\rm i}(\tau_{2}) H_{\rm i}(\tau_{1}) }_{0} 
\label{eq:2017-04-04-11-41} \\
q_{n} =& \{ \tau_{1}, \tau_{2}, \cdots, \tau_{n} \}
\label{eq:2017-04-04-11-42} \\
\int_{n} d\tau =& \dfrac{1}{n!} \int_{0}^{\beta} d\tau_{n} \cdots d\tau_{2} d\tau_{1}.
\label{eq:2017-04-04-11-43} 
\end{align}
In the Monte Carlo method,
the Markov chain of the imaginary time arrangement $q_{n}$ that follows the probability distribution $P(q_{n})$ is generated. 
The statistical average of $A$ is estimated as
\begin{align}
\braket{A}
=&
\dfrac{\sum_{n} \int_{n} d\tau P(q_{n}) \frac{\braket{\hat{P}(q_{n})A(q_{n})}_{0}}{P(q_{n})}}{\sum_{n} \int_{n}d\tau P(q_{n})}
\nonumber \\
\simeq&
\dfrac{1}{N_{\rm s}} \sum_{ \{q_{n} \} } \dfrac{ \braket{\hat{P}(q_{n})A(q_{n})}_{0} }{P(q_{n})}
\label{eq:2017-04-04-11-58}
\end{align}
We can use the Wick's theorem since $H_{0}$ has a quadratic form in terms of the creation and annihilation operators of electrons.
Then we obtain
\begin{align}
P(q_{n})
=&
(-U)^{n} \prod_{\sigma}{\rm det} {\cal G}_{\sigma},
\label{eq:2017-04-04-12-08}
\end{align}
where $({\cal G}_{\sigma})_{ij} = {\cal G}_{\sigma}(\tau_{i}-\tau_{j})$ is the zeroth order Green's function in the impurity model.

\section{Results} \label{sec:Result}

To see the electron-hole asymmetry of the correlation strength, 
 we analyze the filling dependence of the quantity $\beta G_{\bm{k}}(\beta/2)$\cite{PhysRevB.80.245102,PhysRevB.80.045120,PhysRevB.82.155101} defined as 
\begin{align}
\beta G_{\bm{k}}(\beta/2)
=&
\dfrac{1}{2\pi T} \int_{-\infty}^{\infty} \dfrac{A_{\bm{k}}(\omega) d\omega}{\cosh (\omega/2T)},
\label{eq:2017-04-03-19-50} 
\end{align}
where 
$A_{\bm{k}}(\omega) = -\frac{1}{\pi} {\rm Im}G_{\bm{k}}(\omega)$
is the spectrum function.
Namely, 
$\beta G_{\bm{k}}(\beta/2)$ represents the number of states near the Fermi level.
In the rest of this paper,
we sometimes describe this quantity as $\beta G$ for simplicity.

\subsection{Computational Details}\label{sec:2017-04-15-14-58}
First,
we obtain the band structure of ${\rm HgBa_{2}CuO_{4}}$ by performing the first principle calculation using the WIEN2k package\cite{wien2k}. We derive the model Hamiltonian based on the first principles calculation since the realistic shape of the density of states is essential to the present analysis.
We employ the density functional theory (DFT) using the generalized gradient approximation\cite{PhysRevLett.77.3865}.
From this band structure,
we obtain a single-orbital model corresponding to Cu $d_{x^{2}-y^{2}}$ with a maximally localized Wannier basis\cite{Kunes20101888,Mostofi2008685}.
Figure.\ref{fig:2017-04-02-23-40}(b) shows the Fermi surface and the density of states  of this model at half filling.
Since the electronic structure of the electron-doped ${\rm Nd_{2}CuO_{4}}$ is very similar to that of ${\rm HgBa_{2}CuO_{4}}$,\cite{Ogura} we adopt this model not only in the hole doped regime, but also in the electron-doped regime. 
We take into account using DMFT the electron correlation effect 
beyond DFT.
We take the following two approaches in the DMFT calculation. 
One is the DCA calculation with the CT-QMC solver.
This approach enables us to obtain the $\bm{k}$ dependence of the self energy, and to analyze the pseudo gap directly, although its numerical cost is rather large. The other is the single-site DMFT calculation with the IPT solver.
This enables us to analyze a wide range of parameters (the band filling $n$,  the temperature $T$, the interaction strength $U$) due to its very low numerical cost, although $\bm{k}$ dependence of the self energy is neglected. 
In the DCA with CT-QMC, 
we restrict the band filling to the range $0.7 \leq n \leq 1.2$, and fix the temperature at $T/W=0.005$, and the interaction strength at $U/W=1.0$ ($W$ is the band width). 
The band filling range considered here corresponds to the doping concentration where the  pseudo gap is observed experimentally. 
In the single-site DMFT with IPT, 
the calculation is performed in the band filling range $0.2 \leq n \leq 1.7$ with several values of $T$ and $U$,
namely $T/W = 0.05, 0.025, 0.01, 0.005, 0.0025, 0.00125$ and $U/W = 0.005, 1.0, 1.5$.
Our intention for hypothetically taking  different values of $U$ in the single-site DMFT is to see how the electron-hole asymmetry of the electron correlation emerges with and without the Mottness, while assuming the same value of $U$ regardless of the (electron or hole) doping. 
We should also note that the single-site DMFT underestimates the tendency toward Mott transition compared to DCA, so that, roughly speaking, $U/W=1.5$ in the single-site DMFT can be considered as corresponding to $U/W=1.0$ in DCA.

\subsection{$\bm{k}$ Dependence of $\beta G$ by the DCA}

\begin{figure}
\vspace{10pt}
\centering
\includegraphics[width=7.5cm,clip]{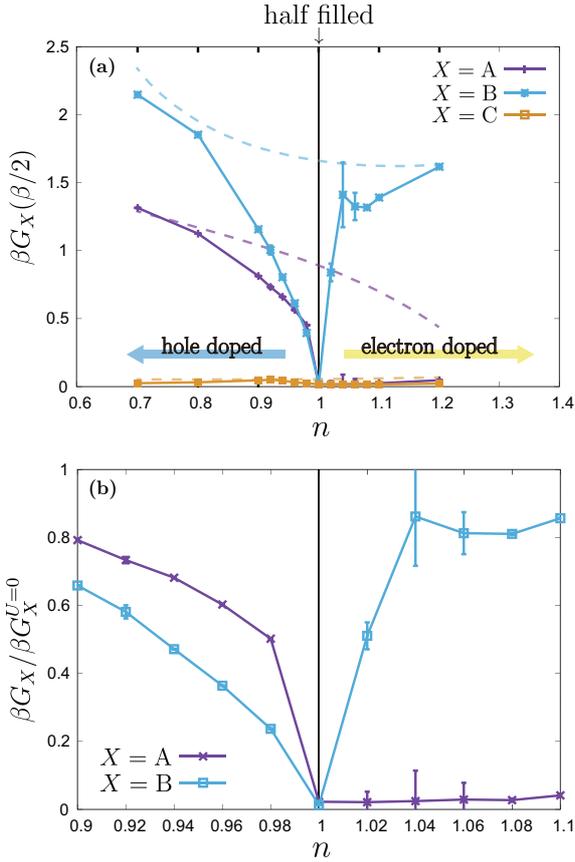} 
\vspace{5pt}
\caption{(Color online) 
(a)
The violet, blue, and orange lines indicate $\beta G_{X}$ defined with Eq. (\ref{eq:2017-04-03-23-30}) of  sectors A, B, and C, respectively. 
The dashed lines indicate $\beta G_{X}$ in the non-interacting case.
(b) The violet and blue lines indicate the ratios $\beta G_{X}/\beta G^{U=0}_{X}$ of  sectors A and B, respectively.
The temperature and interaction strength are $T/W=0.005$ and $U/W=1.0$. 
}
\label{fig:2017-04-03-23-42}
\end{figure}

As mentioned in Sec.\ref{sec:2017-04-15-15-01},
in the DCA calculation,
we divide the Brillouin zone into four sectors as shown in FIG. \ref{fig:2017-04-02-23-40}(a).
$\beta G$ in sector $X$ $(X={\rm A,B,C})$ is defined as 
\begin{align}
\beta G_{X}(\beta/2)
=&
\dfrac{1}{N_{\rm c}}\sum_{\bm{k}\in  X} \beta G_{\bm{k}}(\beta/2),
\label{eq:2017-04-03-23-30}
\end{align}
where $N_{\rm c}$ is the number of $k$ points in the sector.
Figure. \ref{fig:2017-04-03-23-42} (a) shows $\beta G_{X}$ (solid line) and $\beta G_{X}^{U=0}$ (dashed line) in $0.7 < n < 1.2$, where $\beta G_{X}^{U=0}$ represents the  non-interacting $\beta G_{X}$, and (b) shows the ratio $\beta G_{X}/\beta G^{U=0}_{X}$ in the underdoped regime $0.9 < n < 1.1$.
One can see that there is a large difference between the hole- and electron-doped sides.
The most important point is the electron-hole asymmetry of $\beta G_{\rm B}$ (blue line in FIG. \ref{fig:2017-04-03-23-42}).
$\beta G_{\rm B}$ is almost equal to $\beta G_{\rm B}^{U=0}$ at $n=0.7$ and approaches 0 as  $n$ approaches $1$, but  
 as soon as  $n$ exceeds $1$, $\beta G_{\rm B}$ suddenly recovers to about 50\% value of $\beta G_{\rm B}^{U=0}$. 
Namely, in the underdoped regime, 
the correlation effect emerges strongly in the hole-doped side and weakly in the electron-doped side.
This tendency was shown in a previous work\cite{PhysRevB.80.245102}, but there $\beta G$ was plotted as a function of the chemical potential, whereas it is plotted against the electron density here, so that it can be seen more clearly that the electron-hole asymmetry is prominent in the underdoped regime. 

As for $\beta G_{\rm A}$ (violet line in FIG. \ref{fig:2017-04-03-23-42}), 
we also observe an electron-hole asymmetry; in the  electron-doped side in particular, $\beta G_{\rm A}$ is almost zero. This, however, should not be taken as an indication of the strong correlation effect as in sector B  
 of the hole-doped case. In fact, it is known that the antinodal regime of the Fermi surface is strongly affected by the spin fluctuation caused by the $(\pi,\pi)$ nesting, a weak coupling effect, when electrons are doped\cite{doi:10.1143/JPSJ.67.1533}. The $2\times 2$  partitioning of the Brillouin zone adopted in the present study overestimates this effect because the only portion of the Fermi surface included in sector A 
 is the very vicinity of the antinodal point. The overestimation can indeed be confirmed by looking into the results in Ref.[\citen{PhysRevB.80.245102}], where the Brillouin zone was divided into eight sectors. There, the electron-hole asymmetry of the antinodal sector is barely seen.

\subsection{$T$ Dependence of $\beta G$ by the Single-site DMFT}

\begin{figure}
\centering
\includegraphics[width=7.5cm,clip]{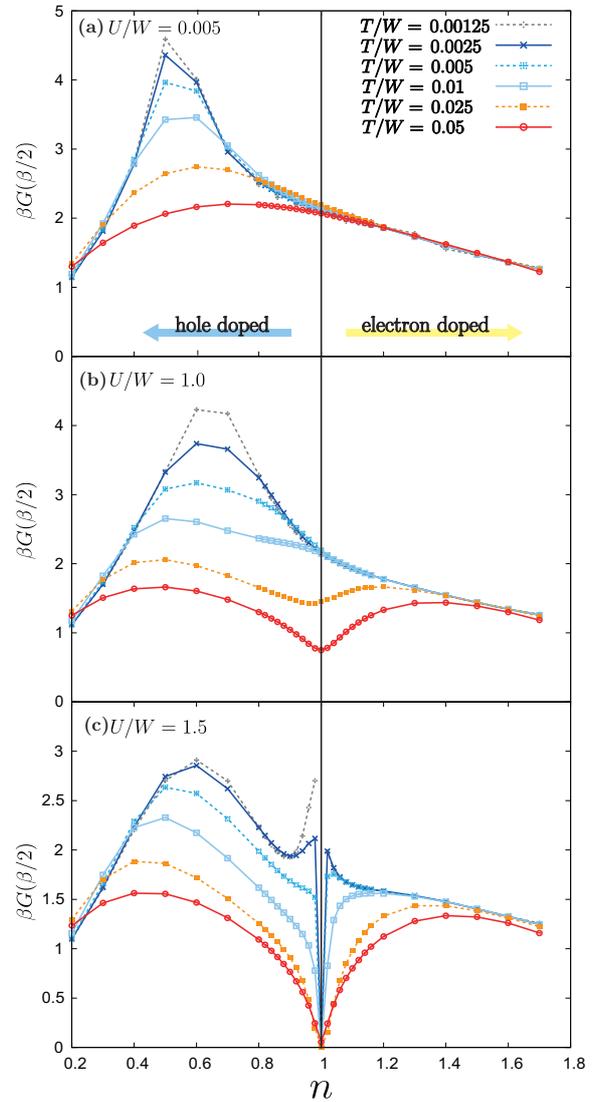} 
\caption{(Color online) 
$\beta G$ 
defined with Eqs. (\ref{eq:2017-04-05-09-38})   
are plotted against the band filling $n$ for 
$U/W=$ (a) 0.005, (b) 1.0, and (c) 1.5 with various temperatures.
}
\label{fig:2017-04-05-01-32}
\end{figure}

We calculate the following average value of $\beta G_{\bm{k}}$ in the Brillouin zone since the self energy is independent of $\bm{k}$ in the single-site DMFT: 
\begin{align}
\beta G(\beta/2) 
=&
\dfrac{1}{N} \sum_{\bm{k} \in {\rm BZ}} \beta G_{\bm{k}}(\beta/2),
\label{eq:2017-04-05-09-38}
\end{align}
where $N$ is the number of $k$ points in the Brillouin zone.
In FIG. \ref{fig:2017-04-05-01-32}(a)-(c),  $\beta G$ is plotted against the band filling $n$ for three values of $U$ and various temperatures.
When $U$ is nearly zero, the temperature dependence of $\beta G$ is barely seen except in the band filling regime where the Fermi level sits close to the van Hove singularity point. When $U$ is increased to an intermediate value where the Mott transition still does not take place at $n=1$, the band filling range where the temperature dependence is present is broadened, 
although the band filling dependence is continuous in this case. A more interesting electron-hole asymmetry is seen in the underdoped regime when $U$ is further increased so that the Mott transition occurs at $n=1$. Namely, the temperature dependence is barely present in the electron-doped regime $(n>1.1)$ when the temperature is low,  as if the electron correlation effect is absent,  while a strong temperature dependence is seen in the hole-doped side\cite{ft_Mills}.  Hence, the electron-hole asymmetry occurs  discontinuously with respect to the band filling 
in the presence of the Mottness. In other words, the ``visibility''  of the electron correlation effect changes abruptly between the hole and the electron-doped cases.

\begin{figure}
\vspace{10pt}
\centering
\includegraphics[width=7.5cm,clip]{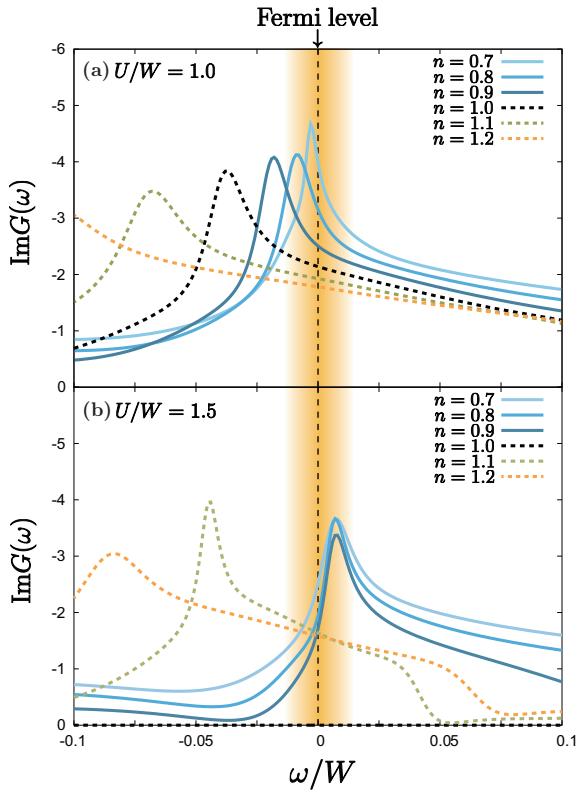} 
\vspace{5pt}
\caption{(Color online) 
The spectrum near the Fermi level: 
The curves with blue tone colors describe the spectrum in the hole-doped side and with yellow tone colors in the electron-doped side.
The black curve describes the spectrum of $n=1$, which vanishes in the lower panel due to the Mottness.
The orange band describes the temperature range for which the integration in $\beta G$ is effective.
The temperature is $T/W=0.005$.

}
\label{fig:2017-04-05-21-04}
\end{figure}

In order to understand the origin of this electron-hole asymmetry,
we investigate the spectrum function $A(\omega)=-\frac{1}{\pi}{\rm Im}G(\omega)$ obtained by performing the analytical continuation with Pad\'{e} approximation\cite{Vidberg1977}.
The enlarged view of the spectrum near the Fermi level at $T/W=0.005$ is shown in FIG. \ref{fig:2017-04-05-21-04}. 
The orange band describes the temperature range for which the integration in $\beta G$ is effective.
In the case of $U/W=1.0$ shown in FIG. \ref{fig:2017-04-05-21-04}(a) (the Mottness is absent), the peak of the spectrum is located in the range $\omega < 0$ reflecting the shape of the non-interacting DOS shown in FIG. \ref{fig:2017-04-02-23-40}(b) that has the van Hove singularity in $\omega<0$. The spectrum moves continuously when the band filling is varied, which leads to the continuous band filling variance of the temperature dependence of $\beta G$ seen in Fig. \ref{fig:2017-04-05-01-32}(b).
The peak of the spectrum in the hole-doped side described by the blue tone colors is closer to the Fermi level than that in the electron-doped side described by the yellow tone colors,
so that a larger temperature dependence exists in the hole-doped side.
By contrast, in the case of $U/W=1.5$ shown in FIG. \ref{fig:2017-04-05-21-04}(b) (the Mottness is present),
the variation of the spectrum with respect to the band filling is qualitatively different from that in $U/W=1.0$.
The peak of the spectrum in the hole-doped side is located within the integration temperature range and barely moves when the band filling is varied,  
while the situation in the electron-doped side is similar to that in the case $U/W=1.0$.
This discontinuous behavior of the spectrum gives rise to the discontinuous temperature dependence shown in FIG. \ref{fig:2017-04-05-01-32}(c).

\section{Conclusion}\label{sec:Conclusion}
In conclusion,
we have shown that the electron-hole asymmetry of the electron correlation effect can exist even under common interaction strengths and the band structure between the hole- and electron-doped systems.  The presence of this electron-hole asymmetry itself is not so surprising because the non-interacting DOS is already asymmetric.
However, the asymmetric feature of the DOS alone cannot account for the striking asymmetry observed in the experiments. Our important finding is that considering the strong correlation effect in addition to the asymmetry of the DOS makes it possible to understand the experimental observation.
To be more precise,  
assuming that the mother compounds of both the electron-doped and the hole-doped cuprates are Mott insulators, the combination of the Mottness and the asymmetry of the DOS can explain the discontinuous electron-hole asymmetry in the phase diagram of the cuprates.
The fact that the present results are obtained under a common value of $U$ between the hole and the electron-doped cases implies that the electron correlation effect is less visible in the electron-doped regime\cite{ft_Gull}.

 In this sense,
it can be said that the origin of the electron-hole asymmetry of the cuprates is the asymmetry of the ``visibility'' of the strong correlation effect.

\section{Acknowledgements}
We are grateful to Yusuke Nomura and Shiro Sakai for providing the CT-INT code, and also providing useful comments regarding the manuscript.
A part of the calculations was performed using supercomputers at the Supercomputer Center, Institute of Solid State Physics, The University of Tokyo.
This study was supported by Grant-in-Aid for Young Scientists (B) (No.JP15K17724),
Grant-in-Aid for Scientific Research on Innovative Areas (No.JP17H05481), Grants-in-Aid for Scientific Research (A) (No.JP26247057) and Grants-in-Aid for Scientific Research (B) (No.JP16H04338) 
from the Japan Society for the Promotion of Science.

\bibliographystyle{jpsj}
\bibliography{reference}

\begin{thebibliography}{10}
\bibitem{ft_Mills}
The tendency that electron-hole asymmetry is barely seen in the high
  temperature regime is consistent with a DMFT analysis given in Refs.
  [\citen{PhysRevB.80.054501,PhysRevB.83.094501}].
\bibitem{ft_Gull}
In Ref. [\citen{PhysRevLett.115.116402}], the superconducting $T_c$ of the
  single band Hubbard model was found to be lower in the hole-doped regime than
  that in the electron-doped regime due to the presence of the pseudo gap,
  assuming the same values of the on-site Hubbard $U$ between them. This
  inconsistency with the experimental observation may be resolved by adopting
  the three band model that explicitly considers the oxygen $p$
  orbitals\cite{Ogura}. Even when the three band model is adopted, the present
  conclusion regarding the electron-hole asymmetry of the electron correlation
  strength will not be affected, since the present conclusion relies only on
  the asymmetric shape of the density of states.
\end{thebibliography}

\end{document}